# Huge Stress-induced Adiabatic Temperature Change in a High-Toughness All-*d*-metal Heusler Alloy

*Rui Cai,*[a, b] *Zhiyang Wei,*[a, b, c, d, *] *Hongjie Ren,*[b, e] *Hanyang Qian,*[b, d] *Xinyu Zhang,*[b] *Yao Liu,*[f] *Xiang Lu,*[b] *Wen Sun,*[b] *Meng Gao,*[b] *Enke Liu,*[d, g] *Jian Liu,*[e] *and Guowei Li*[b, d]

[a] *Dongguan Key Laboratory of Interdisciplinary Science for Advanced Materials and Large-Scale Scientific Facilities, School of Physical Sciences, Great Bay University, Dongguan, 523000, China*
[b] *CAS Key Laboratory of Magnetic Materials and Devices, Ningbo Institute of Materials Technology and Engineering, Chinese Academy of Sciences, Ningbo 315201, China*
[c] *Great Bay Institute for Advanced Study, Dongguan 523000, China*
[d] *University of Chinese Academy of Sciences, Beijing 100049, China*
[e] *School of Materials Science and Engineering, Shanghai University, Shanghai 200444, China*
[f] *State Key Laboratory for Mechanical Behavior of Materials, and Frontier Institute of Science and Technology, Xi'an Jiaotong University, Xi'an 710049, China*
[g] *State Key Laboratory for Magnetism, Beijing National Laboratory for Condensed Matter Physics, Institute of Physics, Chinese Academy of Sciences, Beijing 100190, China*
*Correspondence author. E-mail address: zywei@gbu.edu.cn.

**Keywords**: thermal management, elastocaloric effect, solid state refrigeration, all-*d*-matel Heusler alloys, energy conversion

**Abastract** The elastocaloric effect (eCE), referring to the thermal effect triggered by a uniaxial stress, provides a promising and versatile routine for green and high efficient thermal management. However, current eCE materials generally suffer from relatively low eCE and poor mechanical properties, hindering their practical applications. Here, we report a exceptionally huge eCE with a directly measured adiabatic temperature change of up to 57.2 K in a dual-phase all-*d*-metal Heusler $Mn_{50}Ni_{37.5}Ti_{12.5}$ polycrystalline alloy, revealing an extra contribution to the latent heat during the stress-induced martensitic transformation from B2 to $L1_0$, and breaking the record of adiabatic temperature change for elastocaloric alloys. Moreover, thanks to the combined strengthening effect of *d-d* hybridization and well-dispersed secondary cubic γ phase, the alloy can endure a uniaxial stress up to 1760 MPa. Such an abnormal huge eCE is attributed to the combination of the enhanced entropy change associated with a stress-induced B2 to $L1_0$ martensitic transformation under higher stress, in contrast with the thermally induced B2 to 5-layer modulated structure one, and the high transformation fraction due to the multi-point nucleation facilitated by the γ phase dispersed in the main phase. This work provides insight into making full use of the transformation heat to enhance the caloric effect for high-efficient thermal management systems.

# 1. Introduction

Making good use of electrons supports the modern life, while precisely control heat (thermal management) remains challenging due to the nature of heat, governed by the second law of thermodynamics. However, the importance of thermal management is increasing recognized. On one hand, the miniaturization of electronic equipment requires high-efficient management of heat[1-4]. On the other hand, to optimize the use of waste heat and to improve the efficiency of heat-manage systems, for instance, refrigerators and heat pumps, can significantly reduce carbon emission. Especially, the wide usage of hydrofluorocarbons refrigerant in the traditional vapor compression refrigeration has contributed significantly to the greenhouse effect[5]. Solid-state caloric effects, including magnetocaloric effect[6-8], electrocaloric effect[9-11], barocaloric effect and elastocaloric effect (eCE)[12-15] offer a promising pathway for high-performance thermal management due to their high energy conversion efficiency, zero global warming potential, and feasibility to be scaled down[16-18].

Among these, eCE is particularly promising, as higher magnitude of heat could be yielded by an available driving field, i.e., uniaxial stress[19-21]. Large eCEs are commonly observed in shape memory alloys (SMAs) that undergoes a martensitic transformation (MT)[22-23], for instance, Cu-based SMAs[24-25], NiTi-based SMAs[26-28], and NiMn-based Heusler alloys[29-32]. Practical applications require high eCE performance materials that simultaneously possess a large adiabatic temperature change ($\Delta T_{ad}$), which ensures a high energy conversion efficiency[14, 22], and excellent mechanical properties to ensure the machinability and fatigue life [22, 33]. However, the current materials can hardly meet these two requirements.

Recently, a novel class of all-$d$-metal Heusler NiMnTi(Co) magnetic shape memory alloys were realized by the concept of $d$-$d$ hybridization in transitional $d$-block elements instead of the $p$-$d$ hybridization in traditional Heusler alloys[34-35]. Compared to conventional Heusler alloys, these alloys exhibit larger entropy changes in the vicinity of MT and improved mechanical properties [36] that originated from competition between the weak $d$-$d$ covalent bonds with the metal bonds in the alloy [37-39]. These characteristics make all-$d$-metal Heusler alloys ideal candidates for solid-state refrigeration and active thermal management systems. Various caloric effects including magnetocaloric, elastocaloric and barocaloric effect have been reported in these alloys [40-44]. Notably, colossal eCEs of $\Delta T_{ad}$= -31.5 K and -37.3 K were observed in a B-doped $Ni_{49}Mn_{33}Ti_{18}$ alloy and an oriented $Ni_{49}Mn_{33}Ti_{18}$ alloy, respectively [43, 45]. These huge $\Delta T_{ad}$s are believed to be closely related with large volume change of -1.89 % during MT for that volume change is deemed as a measure of lattice distortion degree[46].

In fact, an higher volume change of -2.54 % was reported in another all-*d*-metal Heusler alloy MnNiTi(Co) series in our previous work[35], which indicates that an even higher eCE could be expected in these alloys. However, the study of eCE for MnNiTi(Co) alloys is still lacking[47-49]. The primary reason is that MT of MnNiTi series is quite sensitive to synthesis condition due to the weak *d-d* hybridization and the heat treatment needs to be carefully optimized to achieve high quality alloys and consequently the high eCE performance.

In this work, polycrystalline alloys of MnNiTi series with reduced hysteresis and transformation temperature span were obtained by optimizing the heat treatment condition [50]. Stress induced MT and eCE were investigated by digital image correlation (DIC) technique and infrared (IR) thermography. Surprisingly, huge $\Delta T_{ads}$ up to 57.2 K were detected upon stress loading, refreshing the record of eCE materials. The origination of such a huge eCE is discussed.

## 2. Experimental

The $Mn_{50}Ni_{37.5}Ti_{12.5}$ (denoted as Ti12.5) alloys were prepared by arc melting high purity metals under an Ar atmosphere. The ingots were re-melted five times for homogenization. Extra Mn of 5‰ was added to compensate Mn volatilization during melting. The samples to be compressed were cut by wire electrical discharge machining into cuboids with a dimension of $4 \times 4 \times 8$ mm$^3$ from the ingot, then sealed in a quartz tube under vacuum and annealed in a tube furnace at 1188 K for 4 days, followed by quenching into ice water. As a reference, $Mn_{50}Ni_{37.5}Ti_{12.5}$ (denoted as Ti11.5) with higher MT temperatures was prepared following the same method as the Ti12.5 alloy.

X-ray diffraction (XRD; D8 ADVANCE DAVINCI) and electron backscatter diffraction (EBSD; GeminiSEM 300) were used to analysis the texture evolution and crystal structure of the alloy, respectively. ATEX software is used for EBSD processing. The TEM sample was fabricated by the focused ion beam. The microstructure was characterized using a transmission electron microscope (TEM; 200 kV JEOL 2100F TEM). Thermo-magnetization curves were measured using a superconducting quantum interference device magnetometer (MPMS; Quantum Design). The cooling and heating rates were set as 3 K min$^{-1}$. Characteristic temperatures of MT and entropy change ($\Delta S$) were determined using differential scanning calorimetry (DSC; NETZSCH DSC 214) with heating/cooling rate of 10 K min$^{-1}$. The specific heat capacity ($C_p$) measurement was conducted by the DSC and calibrated using a standard sapphire sample.

The eCE test was carried out on a universal experimental machine (Zwick/RoellZ100) at an ambient temperature of 270 K. A combination of in situ DIC (Correlation Solutions Inc.)

and IR thermography (FLIR A325sc) was utilized to monitor the evolution of two-dimensional strain and the temperature of the sample during the stress loading/unloading. A loading/unloading rate of 500 mm min$^{-1}$ was adopted to ensure the quasi-adiabatic conditions during eCE measurement. The strain in the compression process is controlled by the displacement of the universal testing machine and the strain in the stress-strain curve is derived from DIC data.

## 3. Results and Discussion

$Mn_{50}Ni_{37.5}Ti_{12.5}$ (denoted as Ti12.5) alloys were synthesized by arc melting of high-purity Ni, Mn, and Ti metals and heat treated by the optimized condition (see Ref.50). According to the XRD pattern in Figure 1a, the Ti12.5 alloy is crystallized in a B2 ordered structure with space group No. 221[35] and a lattice parameter $a_{B2}$ = 2.974 Å, indicating it is in austenitic state at room temperature (RT). The weak peaks between the austenitic peaks are observed and indexed as cubic γ phase (the actual composition is $Mn_{67.9}Ni_{28}Ti_{4.1}$ determined by Energy dispersive spectrometer) with space group No. 225 and a lattice parameter of $a_γ$ = 3.710 Å. The microstructure of Ti12.5 was further studied by TEM. Figure 1b shows TEM image of Ti12.5 that contains B2 austenite and γ phase and their phase boundary as indicated by the red dashed line. High-magnification image of B2 austenite along the [111]$_{B2}$ zone axes and γ phase along the [110]$_γ$ zone axes, and corresponding selected area electronic diffraction (SAED) patterns are presented in Figure 1c. in Figure 1c the SAED pattern along the [001]$_{B2}$ zone axes, as shown in the top half Figure 1d. Accordingly, the lattice parameter can be roughly determined as $a_{B2}$ = 2.976 Å for B2 austenite and $a_γ$ = 3.631 Å for the γ phase, is almost consistent with XRD results. From the EBSD image in Figure 1d, Ti12.5 alloy consists of the dominated austenitic phase with size of 100 ~200 μm as illustrated by the inverse pole figure, and a small amount of γ phase with diameters of around 10 μm (in navy blue) that dispersed within the grain and grain boundary of main phase. Such a microstructure is confirmed over large areas of the alloy by scanning electron microscope (SEM) images in the Figure S1. It is needed to be pointed out that the EBSD image only reflects the area proportion of different phases of the sample in the scanning area, rather than the phase proportion. In fact, the γ phase takes only a small proportion, evidenced by its low peak intensity in the XRD pattern. The ductile γ phase disperses in the main phase is expected to lower the energy barrier of MT [33], and is reported to be able to improve the mechanical properties of the alloy[51-53].

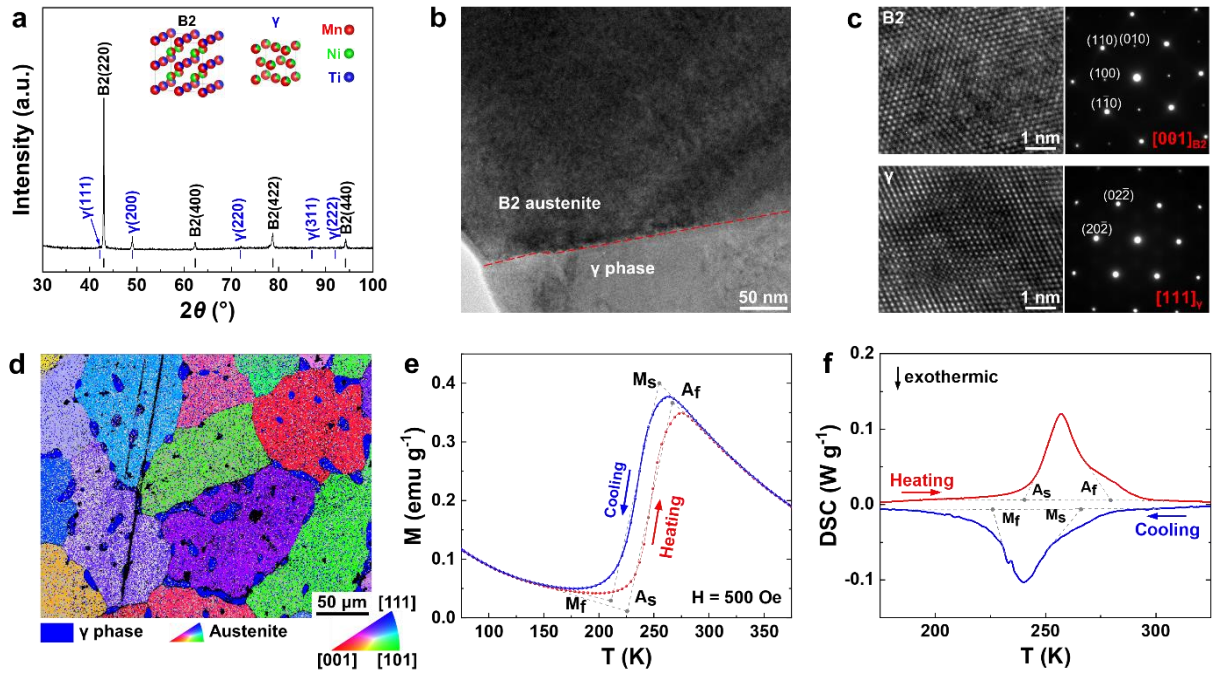

**Figure 1.** a) RT XRD pattern of Ti12.5 alloy and schematic crystal structure of B2 austenite and γ phase. b) The bright-field TEM image of the B2 austenite and γ phase of Ti12.5 alloy. c) The high-magnification image of (b) and corresponding SAED pattern along the $[001]_{B2}$ zone axes of B2 austenite, and the $[111]_γ$ zone axes of γ phase, respectivley. d) EBSD image of the alloy, indexed according to the lattice parameters obtained by XRD. The navy blue represents the γ phase, and remaining colored regions represent orientation distributions of B2 austenite. e) *M-T* curves of the Ti12.5 alloy in a magnetic field of *H* = 500 Oe. f) DSC curves of the Ti12.5 alloy upon heating and cooling with a temperature changing rate of 10 K min$^{-1}$.

Figure 1e presents the thermo-magnetization (*M-T*) curve of the Ti12.5 polycrystalline alloy in a magnetic field of *H* = 500 Oe. The Ti12.5 alloy is in low-magnetization state within the measured temperature range, consistent with the previous results [34-35]. The magnetization jumps at upon cooling and heating with a thermal hysteresis suggests the occurrence of the first order MT and inverse MT. The characteristic temperatures of MT determined from *M-T* curves are listed in Table S1, and they are all below RT, in agreement with the XRD result. Notably, the thermal hysteresis of the sample here is reduced to be $\Delta T_{hys}$ = 13.0 K, which benefit the energy conversion efficiency [21]. The MT and inverse MT are also confirmed by the exothermic and endothermic peaks in differential scanning calorimetry (DSC) curves, as shown in Figure 1f (see Table S1 for MT characteristic temperatures). By integrating the exothermic peak of DSC cooling curve, combined with the $C_p$ data (Figure S2), the thermal-induced MT entropy change and the adiabatic temperature change could be estimated as large as $\Delta S$ = 65.8 J kg$^{-1}$ K$^{-}$

$^1$, and $\Delta T_{ad} = T\Delta S/C_p$ = 39.9 K, respectively. Such values suggest a large eCE potential in the alloy.

The stress-induced MT behaviors and corresponding elastocaloric effect of Ti12.5 alloy were studied by *in situ* methods of IR thermography and DIC strain measurement. Figure 2b shows the stress-strain curve of Ti12.5 alloy measured under an eCE cycle with a large applied overall strain of 13%, and corresponding DIC and IR images at various stages of the cycle are displayed in Figure 2c. The non-homogenous strain was observed in the early stage of the loading process (II) from the DIC images, in contrast with that of all-*d*-metal Heusler NiMnTi polycrystalline alloy where the strain develops in one region and then spreads out to other regions [37-38]. This indicates the MT of Ti12.5 polycrystalline nucleates and grows in many regions of the sample with increasing the overall strain (This feature is more evident at a lower strain level as illustrated in DIC image in Figure 3d). The possible reason accounting for this multiple nucleation of MT may rely in the dispersed secondary γ phase serving as nucleation site [33, 54]. At stage IV, the strain distribution tended to be homogeneous, within a narrow range from 11% to 14%, implying a uniform MT over the sample. Further increasing the overall strain, the sample exhibit a strain concentration behavior, with local strains up to 20% at the areas where MT occurs preferentially in the early stage of the loading. Approaching the end of loading (V ~ VI), the slope of the stress-strain curve rose, suggesting the MT nearly completed. From the DIC images (V ~ VI), the sample tended to bend and the high local strain area did not fully recover after the cycle, indicating plastic deformations occurred.

Notably, the corresponding maximum stress is up to 1.53 GPa without fracture and it even bent, exhibiting an excellent toughness and machinability of Ti12.5 alloy. To evaluate the limit of the alloy, a fresh sample was subjected to a high stress until fracture. The stress-strain curve is displayed in Figure 2a. The alloy exhibited a typical high toughness of 452.5 ± 23.5 J cm$^{-3}$ and a high facture strength of ~1.76 GPa with a high ductility with a compression rate to over 32%. Such a high strength is higher than NiMnTi all-*d*-metal Heusler alloys[37-38, 43, 45, 55], and other elastocaloric materials, as shown in Figure 2e and Table S2. Both experimental and theoretical results confirmed the good mechanical properties in all-*d*-metal Heusler alloys[35, 36, 43, 56-57]. The underlying mechanism of high toughness of these alloys lies in the competition of metallic bond with the relatively weak *d-d* hybridization[37, 39]. As for our dual-phase Ti12.5, the well-dispersed ductile γ phase in the main Heusler phase further strengthens the alloy[51-53]. These two aspects result in a synergic effect and endue our alloys with a high toughness.

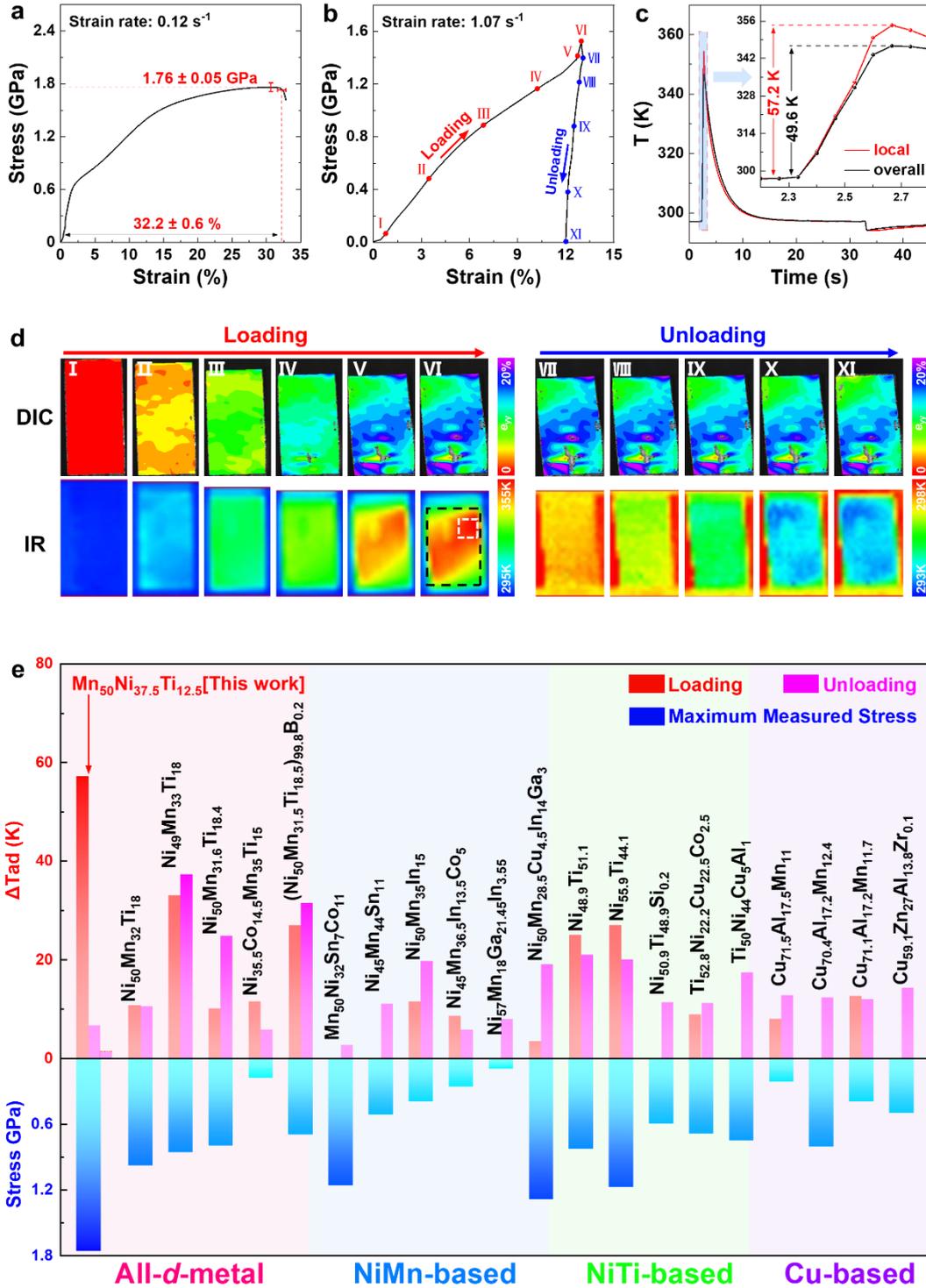

**Figure 2.** a) The RT strain-stress curve for Ti12.5 sample with a loading rate of 0.12 s$^{-1}$. b) The RT strain-stress curve for Ti12.5 sample with a loading/unloading rate of 1.07 s$^{-1}$. c) The temperature-time curves correspond to the local area marked by the white dashed rectangle and overall average marked by the black dashed area in (d). The inset is an enlarged view of the temperature time curve of the loading process. d) DIC strain contour images and IR images of the corresponding points in the stress-strain curve in (b). The $e_{yy}$ represents the strain in the compression direction of the sample. e) Comparison on the maximum stress and $\Delta T_{ad}$ between the $Mn_{50}Ni_{37.5}Ti_{12.5}$ alloy and some other elastocaloric materials reported in the literatures (see Table S3).

Now we focus on the elastocaloric $\Delta T_{ad}$ of the Ti12.5 sample. As illustrated by the IR images in Figure 2d, the sample's temperature rose homogeneously during the initial stages of the loading (I~IV), indicating a multisite nucleatation of martensite. This finding aligns with the well-dispersed strain distribution observed in the DIC results. Notably, near the end of loading (IV~V), local areas with higher temperatures began to emerge (marked in white dash box). Analysis of the IR images reveal a delay in reaching the local maximum temperature compared to the overall temperature, as depicted in Figure S3 (with corresponding discussion in Supplementary materials). This delay in maximum local $\Delta T_{ad}$ region suggests an additional transformation contributing to the abnormally high $\Delta T_{ad}$ in these specific areas. Uneven distribution of $\Delta T_{ad}$ was also observed during stress unloading (VII~XI). Figure 2c shows the derived temperature-time curves from IR measurement. Strikingly, a maximum local $\Delta T_{ad}$ as high as 57.2 K (white dashed box in Figure 2d) and an overall average $\Delta T_{ad}$ of 49.6 K (black dashed box) were observed. This exceptional eCE was successfully reproduced in another Ti12.5 sample ($\Delta T_{ad}$ = 64.5 K, Figure S4, Supplementary materials). These huge $\Delta T_{ad}$s surpass all the reported eCE alloys so far and refresh the record of direct measured $\Delta T_{ad}$ (-37.3 K for $Ni_{50}Mn_{32}Ti_{18}$) of eCE materials [45]. As shown in Figure 2e, our Ti12.5 alloy exhibits not only excellent mechanical properties but also the highest $\Delta T_{ad}$ over other typical elastocaloric alloys including NiMn-based Heusler alloys, NiTi alloys, Cu-based alloys and even all-*d*-metal Heusler NiMnTi alloys. Note that the irreversibility of the $\Delta T_{ad}$ is considerable and the $\Delta T_{ad}$ upon unloading is only -3 K. Such an irreversibility of eCE is attributed to the plastic deformation under the large applied stress, as confirmed by the large local strain (up to 20%) in the DIC images.

In fact, at lower applied strains, the reversibility of eCE for Ti12.5 has been improved. Figure 3a illustrates the overall and maximum local $\Delta T_{ad}$ derived from the IR measurement in different loading/unloading cycles for another Ti12.5 sample. For the third eCE cycle (displacement of 1.0 mm), the overall average $\Delta T_{ad}$ = 9.7 K (marked in white dash box) upon loading is slightly higher than the value 6.6 K upon unloading. As the applied strain increases, both the overall and the maximum local $\Delta T_{ad}$ gradually increased during stress loading and such a tendency is more pronounced for the maximum local $\Delta T_{ad}$, as shown in Figure 3b. The irreversibility of $\Delta T_{ad}$, indicated by the difference between absolute value of $\Delta T_{ad}$ upon unloading and $\Delta T_{ad}$ upon loading, increases with higher applied strains. Similar behavior was observed in non-oriented polycrystalline $Ni_{50}Mn_{32}Ti_{18}$ alloy in our previous study [37]. This is a common phenomenon in polycrystalline alloys due to the binding force is weak at the grain boundary and the different forces of grains with different orientations. Therefore, the region where the local maximum $\Delta T_{ad}$ occurs in different eCE cycles is different (so it is not marked in Figure 3d). Despite this, the maximum local $\Delta T_{ad}$ reached 35.6 K during the last compression cycle, approaching the theoretical value of $\Delta T_{ad} = 39.9$ K. As depicted by the IR map of the third cycle in Figure 3d, the inhomogeneity feature of $\Delta T_{ad}$ (stage VI) indicates an incomplete MT of the sample.

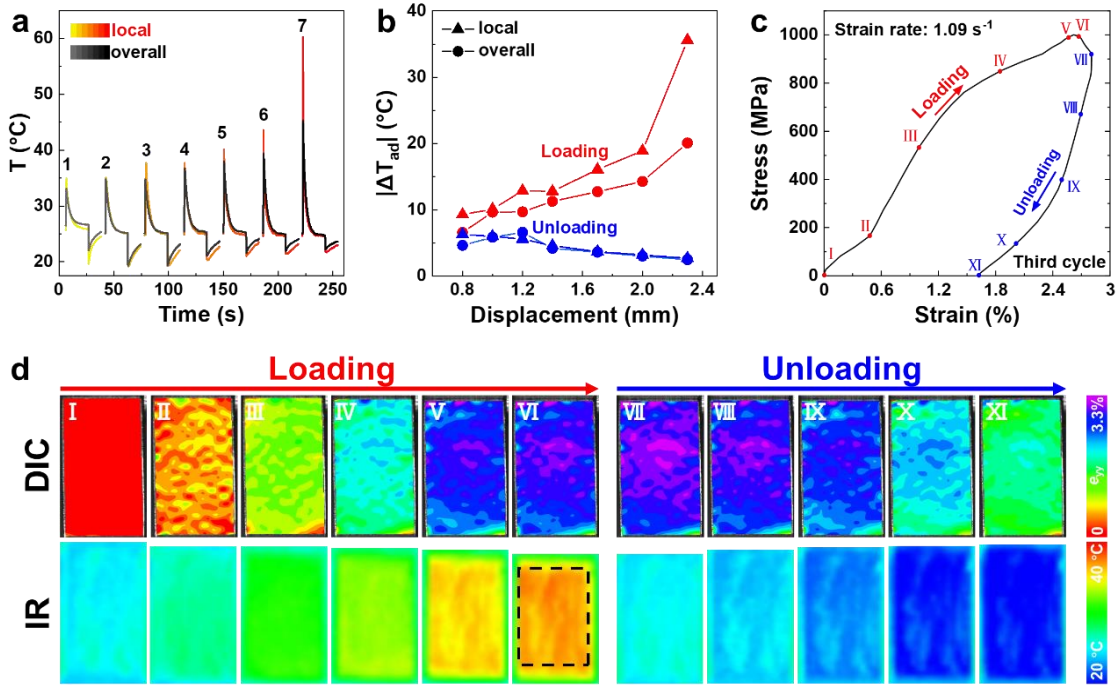

**Figure 3.** a) The temperature-time curves derived from the IR measurement for the Ti12.5 sample under various strain loading cycles with a loading/unloading rate of 500 mm min$^{-1}$. For each cycle, the load was hold for 30 s before unloading. The yellow-red curves represent the local maximum $\Delta T_{ad}$ of the sample that local region area is same as Figure 2d, and the gray-black curves represent the average $\Delta T_{ad}$ of the overall sample marked by the black dashed rectangle. b) The loading displacement dependent $\Delta T_{ad}$ for different loading/unloading cycles. c) The room-temperature strain-stress curve for Ti12.5 sample with a loading/unloading rate of 1.09 s$^{-1}$ corresponding to the third loading cycle. d) DIC strain contour images and IR images of the corresponding points in the stress-strain curve in (c). The $e_{yy}$ represents the strain in the compression direction of the sample.

To elucidate the irreversibility of eCE, DIC and IR measurements were conducted on the third loading cycle, as illustrated in Figure 3d. At the beginning of compression (II~III), strain concentration regions were uniformly distributed, indicating multiple nucleation of MT that induced by the dispersed secondary γ phase[33]. The strain concentration is typical for polycrystalline alloys where varied grain orientations necessitates different critical stresses needed to drive the MT[57-60]. The obvious temperature rising observed from IR image (II~III) also confirms the onset of the MT. As the loading progressed (IV~VI), strain concentrate became more pronounced, as indicated by the purple region in the DIC image (V). $\Delta T_{ad}$ ceased to increase once strain reached its maximum for the cycle. After the cycle, residual strain (in light blue region (XI)) was predominantly concentrated in the area that experiences significant strain during loading (in light purple region (F)), indicating a plastic deformation. This should account for the irreversibility of the eCE. Despite the weak *d-d* covalent bonds[37] inherent in this sort of alloys, the plastic deformations of Ti12.5 alloy primarily arises from the intergranular slip in non-oriented polycrystalline sample, with the γ phase further enhancing the effect by providing more slip direction[61-62]. Such plastic deformations are expected to be overcome by grain engineering method, for instance, directional solidification, and the associated work is ongoing.

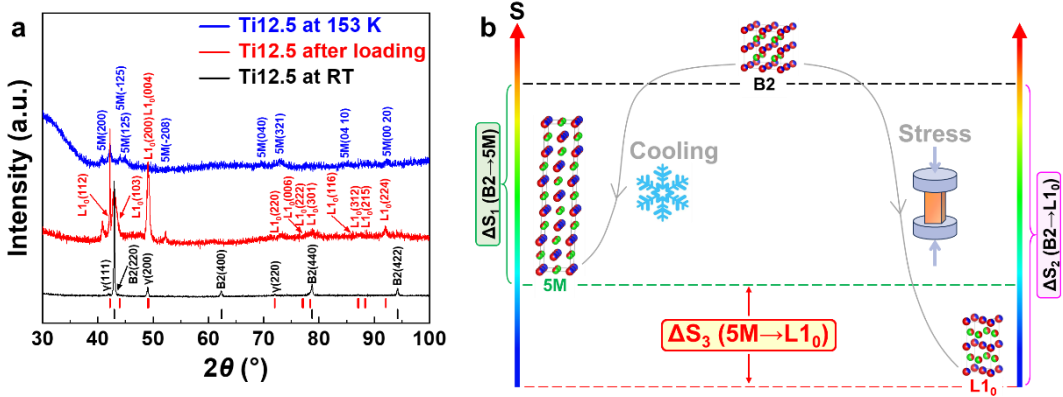

**Figure 4.** a) RT XRD patterns of Ti12.5 alloy before and after the large strain compression (corresponding to the sample in Figure 2b), and XRD pattern of Ti12.5 alloy at 153 K. b) Schematic of the thermally induced/stress-induced MT and the corresponding different product phases with different entropy changes.

Now, we focus on the underlying mechanism of the exeptional eCE. Both the local $\Delta T_{ad}$ = 57.2 K and overall $\Delta T_{ad}$ = 49.6 K upon loading process are higher than the theoretical maximum $\Delta T_{ad}$ = 39.9 K estimated by the DSC measurement. From the measured $\Delta T_{ad}$ = 57.2 K, the corresponding entropy change is calculated as $\Delta S = C_p \Delta T_{ad}/T$ = 94.4 J kg$^{-1}$ K$^{-1}$. Taking into account the non-adiabatic conditions, it is much higher than the $\Delta S$ = 65.8 J kg$^{-1}$ K$^{-1}$ measured by DSC (Figure 1f). To reveal the origin of this abnormal $\Delta T_{ad}$, we examined the product of the stress-induced MT. Figure 4a shows XRD patterns of the Ti12.5 sample before (in black) and after (in red) high stress loading. Different from the thermally induced martensite with a 5-layer modulated (5M) modulated martensite (recorded at 153 K), the stress-induced martensite is dominated by tetragonal L1$_0$ martensite for Ti12.5, and only a small amount of 5M martensite remains. The L1$_0$ tetragonal structure belongs to space group No. 139, and the lattice parameters are defined as $a = b = 3.7054$ Å and $c = 7.4235$ Å. In terms of thermodynamics, the modulated martensite is believed to be a metastable intermediate phase while the L1$_0$ martensite is a thermodynamic stable state with lower free energy[63-64]. For clarity, a schematic is given in Figure 4b: the stress-induced $\Delta S_2$ corresponding the MT from B2 to L1$_0$, is higher than $\Delta S_1$ corresponding to the thermally induced MT from B2 to 5M, giving a difference value of $\Delta S_3$ (5M to L1$_0$). In other words, the stress-induced transformation from austenite to L1$_0$ martensite will release more latent heat than the transformation to 5M modulated martensite. Therefore, stress-induced MT would boost the eCE and thus yield a higher $\Delta T_{ad}$, which is the fundamental reason why the $\Delta T_{ad}$ of the sample is higher than the theoretical value. This is also evidenced by the large volume change up to -3.13% associated with the stress-induced MT, calculated from the lattice parameters of austenite and L1$_0$ martensite.

To evaluate the contribution of the 5M to L1$_0$ MT, a neighbor composition of Mn$_{50}$Ni$_{38.5}$Ti$_{11.5}$ (Ti11.5) in its martensite at RT (Figure S5, Supplementary materials) was selected and subjected to a similar high strain as that of Ti12.5 alloy (Figure 2a). The contribution of 5M to L1$_0$ MT was estimated about 18.0 K (Figure S6, Supplementary materials). That is to say, the theoretical contribution of MT to $\Delta T_{ad}$, should be theoretical $\Delta T_{ad}$ = 39.9 K of B2 to 5M plus actually measured $\Delta T_{ad}$ = 18.0 K 5M to L1$_0$, namely 57.9 K (note that the eCE test is only quasi-adiabatic). In addition, for thermoelastic martensitic materials, the contribution of frication heat to the $\Delta T_{ad}$, including the stress-induced MT and plastic deformation, should be considered especially under fast loading[65-66]. However, the local maximum $\Delta T_{ad}$ doesn't occur in the region where with the biggest strain, but in the region where the shape of sample is basically intact (Figure S4, Supplementary materials). This further verifies the extra $\Delta T_{ad}$ must be attributed to the stress-induced B2 to L1$_0$ MT, as discussed above.

## 4. Conclusion

In conclusion, an all-*d*-metal Heusler Mn$_{50}$Ni$_{37.5}$Ti$_{12.5}$ polycrystalline alloy with reduced hysteresis and large transformation entropy change is obtained by appropriate heat treatment condition. The dual-phase microstructure of the alloy with secondary cubic γ phase dispersed in and/or between the grains of the main combined with the intrinsic ductile nature from the *d*-*d* hybridization of all-*d*-metal Heusler alloy, exhibiting a toughness of 452.5 ± 23.5 J cm$^{-3}$ and a facture strength of ~1.76 GPa. Crucially, supported by the excellent mechanical properties, an unprecedented elastocaloric effect with a local maximum $\Delta T_{ad}$ up to 57.2 K, and an average overall $\Delta T_{ad}$ up to 49.6 K were directly measured, refreshing the record of eCE materials. Such a huge elastocaloric effect is ascribed to the stress-induced B2 austenite to tetragonal $L1_0$ MT that yields more latent heat than the thermally induced B2 to 5M MT. The *in situ* DIC strain measurement and IR thermography reveals the multi-point nucleation initiated by the secondary γ phase efficiently leads to a high transformation fraction and thus further enhance $\Delta T_{ad}$. These results indicate the all-*d*-metal Heusler alloy MnNiTi alloys are promising materials for high performance elastocaloric thermal management, and provide a strategy to unlock the caloric potential for all-*d*-metal Heusler alloys.

**Acknowledgements**

This work was financially supported by Ningbo Natural Science Foundation (Grant No. 2022J292), the National Natural Science Foundation of China (Grant No. 52271194, 52371192, 52088101, 52201234), the State Key Development Program for Basic Research of China (No. 2019YFA0704900), Ningbo Yongjiang Talent Introduction Programme (2022A-090-G), and the Hundred Talents Programs in the Chinese Academy of Sciences. Zhiyang Wei acknowledge the support from the Starting Fund for Talents of Great Bay University. Guowei Li thanks the support from the Max Planck Partner Group program.

**Supplementary materials**

Supplementary material associated with this article can be found, in the online version, at xxx

# Supplementary materials

**Analysis of microstructure for Ti12.5 alloy.**

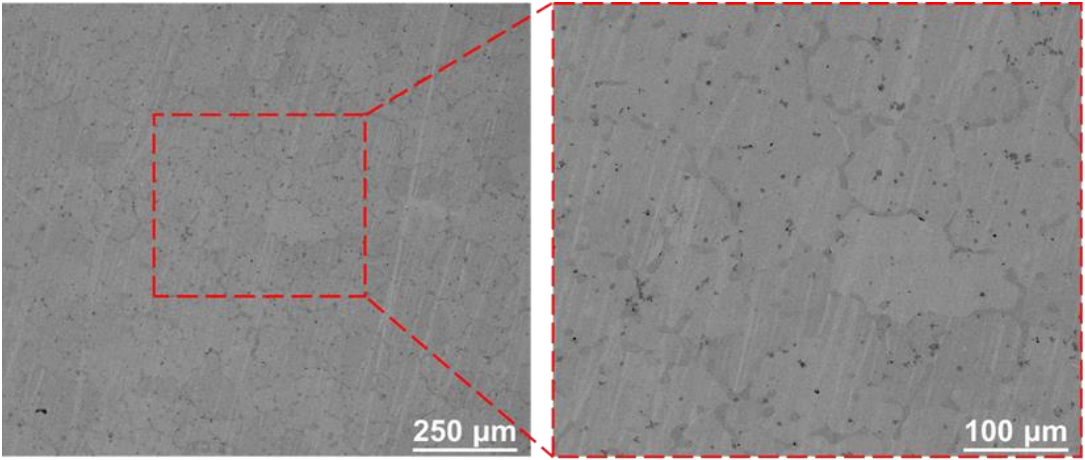

**Figure S1.** The SEM image of Ti12.5 alloy.

**Transformation entropy change and theoretical adiabatic temperature change for Ti12.5 alloy**

From the DSC curves in Figure S1, the transformation entropy change could be estimated as $\Delta S = \Delta H/T = 65.8$ J kg$^{-1}$ K$^{-1}$, where $T = T_M = 240.1$ K. Thus, the $\Delta T_{ad}$ could be estimated as $\Delta T_{ad} = -T\Delta S / C_p = 39.9$ K, where heat capacity $C_p = 490$ J kg$^{-1}$ K$^{-1}$, and the $C_p$ is taken as the average of the specific heat capacity of martensite and austenite at the ambient temperature $T = 297$ K, as seen in Figure S2.

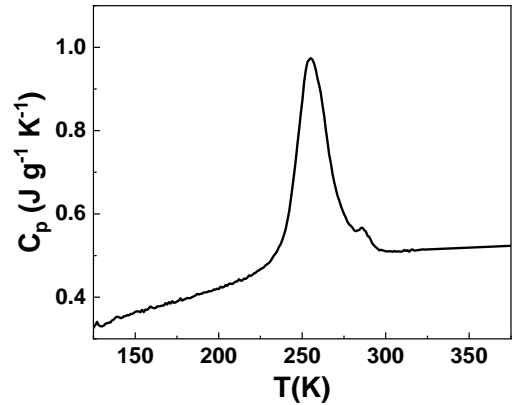

**Figure S2.** The temperature dependence of the specific heat capacity measured upon heating for Ti12.5 alloy.

**Table S1.** The characteristic temperatures of martensitic/austenitic transformation for the Ti12.5 alloy determined from magnetic and DSC measurements, using the tangential method. $M_s$, $M_f$ and $T_M$ are the start, finish and peak temperature (at which the slope of the *M-T* curve is maximum) of MT, and $A_s$, $A_f$ and $T_A$ are start, finish, and peak temperature of austenitic transformation, respectively. The thermal hysteresis ($\Delta T_{hys}$) of the sample is taken as $\Delta T_{hys} = T_A - T_M$.

|     | $M_s$ (K) | $M_f$ (K) | $A_s$ (K) | $A_f$ (K) | $T_M$ (K) | $T_A$ (K) | $\Delta T_{hys}$ (K) |
| --- | --- | --- | --- | --- | --- | --- | --- |
| M-T | 255.1 | 210.8 | 225.7 | 266.7 | 235.0 | 248.0 | 13.0 |
| DSC | 266.0 | 226.1 | 240.7 | 279.0 | 240.1 | 257.1 | 17.0 |

**Analysis of temperature-time curve for Ti12.5 alloy.**

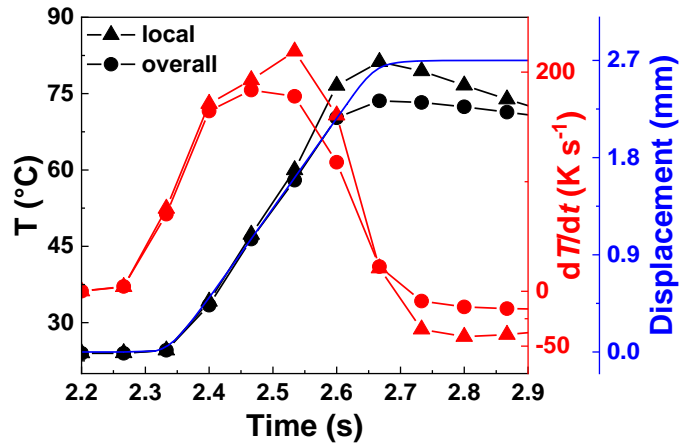

**Figure S3.** The derivatives of the temperature-time curves in Figure 2b during stress loading. The displacement-time curve is also plotted as a reference.

In order to explore the source of the extra heat, the derivatives of temperature-time curves during loading in Figure 3b are given in Figure S3. Since the $\Delta T_{ad}$ of the sample is mainly supplied by MT, the derivative of the temperature-time curves should reflect the transformed martensite proportion per unit time in the sample. Upon loading, the derivatives of both overall and maximum local $\Delta T_{ad}$ increased at first, indicating the MT speeding up, and then reach a maximum value at $t = 2.47$ s for the overall $\Delta T_{ad}$ and $t = 2.53$ s for the maximum local $\Delta T_{ad}$. Such a delay in maximum local $\Delta T_{ad}$ region points to another transformation that should be accounting for the abnormal high $\Delta T_{ad}$ in this local region.

**Repeated experiment on huge elastocaloric effect for Ti12.5 alloy (a fresh sample).**

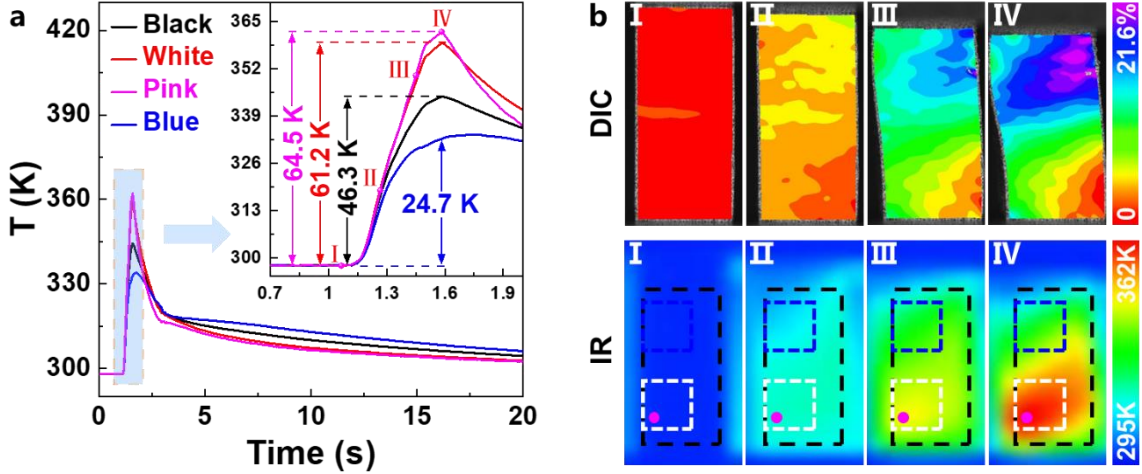

**Figure S4.** a) The temperature-time curves correspond to different area in (b). The inset is an enlarged view of the temperature time curve of the loading process. b) DIC strain contour image and IR images of the corresponding points in the temperature-time curve in (a).

The huge eCE was successfully reproduced in another Ti12.5 sample, its temperature-time curves correspond to different area as shown in Figure S4a. Figure S4b shows the corresponding DIC strain contour image and IR images of the corresponding points in the temperature-time curve in Figure S4a. it can be seen that serious plastic deformation occurred in the upper part of the sample, but the temperature change of the sample was mainly concentrated in the lower part of the sample. That just goes to show the contribution of friction heat to $\Delta T_{ad}$ is much smaller than heat of martensitic transformation.

**Elastocaloric effect of Ti11.5 in martensite state**

Figure. S5a shows the *M-T* curves for $Mn_{50}Ni_{38.5}Ti_{11.5}$ (Ti11.5) alloys. Thus, Ti11.5 was selected to determine the crystal structure of thermally induced martensite, and to be tested for eCE at room temperature to help analyze the extra heat during the stress-induced MT for Ti12.5 alloy. It is observed Ti11.5 undergoes a martensitic transformation above room temperature. From the room temperature XRD pattern in Figure S5b, the Ti11.5 alloy is almost in 5-layer modulated (5M) martensite state at room temperature with lattice parameters of *a* = 4.4285 Å, *b* = 5.402 Å, *c* = 21.4082 Å and *β* = 92.892°.

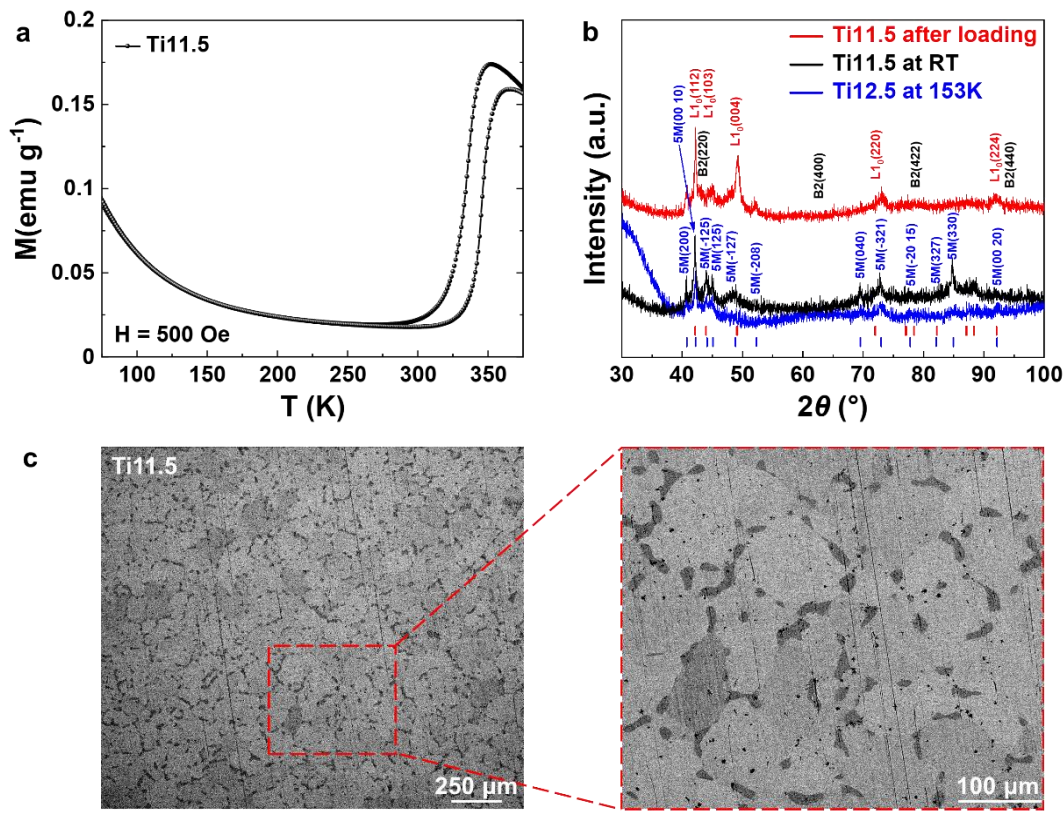

**Figure S5.** (a) *M-T* curves of the Ti11.5 alloy in a magnetic field of 500 Oe. (b) Room temperature XRD patterns of Ti11.5 alloy before and after loading and XRD patterns of Ti12.5 alloy at 153 K. (c) SEM images of Ti11.5 alloy.

Figure S5c shows SEM images of Ti11.5 alloy. The microstructure of Ti11.5 samples is similar to Ti12.5 sample (Figure S1) and the corresproding atomic ratio of austenite and γ phase in Ti12.5 and Ti11.5 samples is obtained by EDS, as shown in Table S2. Similarly, Ti11.5 was subjected to a fast stress loading, and XRD pattern of the alloy after loading (Figure 4a and Figure S5b) suggests it is composed of $L1_0$ martensite main phase and a small amount of 5M martensite, similar to that for Ti12.5 alloy.

**Table S2.** Atomic ratio of austenite and γ phase in Ti12.5 and Ti11.5 samples obtained by EDS.

| Phase | Mn (%) | Ni (%) | Ti (%) |
|---|---|---|---|
| austenite from Ti12.5 | 45.70 ± 0.26 | 40.83 ± 0.26 | 13.47 ± 0.19 |
| austenite from Ti11.5 | 47.76 ± 0.36 | 40.17 ± 0.35 | 12.07 ± 0.27 |
| γ phase from Ti12.5 | 67.72 ± 0.18 | 28.17 ± 0.05 | 4.11 ± 0.16 |
| γ phase from Ti11.5 | 67.47 ± 0.33 | 28.50 ± 0.41 | 4.03 ± 0.20 |

The stress-strain curve (calibrated according to the length of the sample after compression) and corresponding temperature-time curve (by infrared thermography) were recorded, as shown in Figure S11. The maximum local adiabatic temperature change ($\Delta T_{ad}$) upon stress loading and unloading for Ti11.5 were $\Delta T_{ad}^{loading}$ = 18.0 K and $\Delta T_{ad}^{unloading}$ = -1.3 K (Figure S11b). The irreversible $|\Delta T_{ad}^{irrev}| = |\Delta T_{ad}^{loading}|-|\Delta T_{ad}^{unloading}|$ =16.7 K should consist of two contributions: the MT from 5M to L1$_0$, friction associated with plastic deformation and MT. Since the enclosed area of the stress-strain curve is proportional to the internal friction and the area (99.2 MPa, Figure S6a) for Ti11.5 is close but slightly higher than that (93.7 MPa, Figure 2a) of Ti12.5[7-8], it reasonable to speculate that for Ti12.5 the friction contribution to the direct measured $\Delta T_{ad}^{fri}$ should be lower than $|\Delta T_{ad}^{irrev}|$ = 16.7 K of Ti11.5.

Note that the uneven temperature distribution in the sample can be clearly seen in Figure. S6c III and IV, indicating that MT is likely to occur in those red regions. The temperature drop in the unloading process on the temperature time curve also evidences the presence of inverse MT in the sample. According to the results in Figure S10b and Figure 3b, such stress-induced MT is ascribed to be from 5M martensite to L1$_0$ martensite. That is to say, there is a considerable contribution of MT to $|\Delta T_{ad}^{irrev}|$ = 16.7 K, and subtracting this contribution, the contribution of friction should be much lower than $|\Delta T_{ad}^{irrev}|$ = 16.7 K.

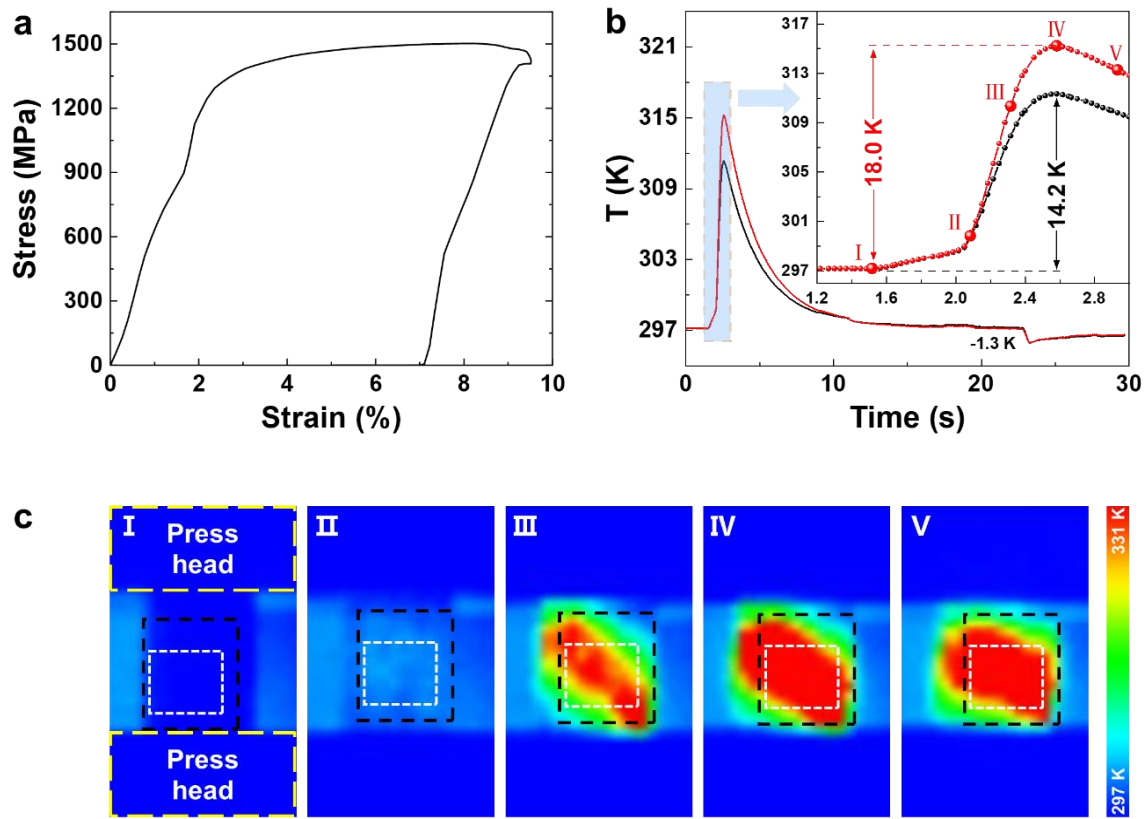

**Figure S6.** (a) Room-temperature strain-stress curves for Ti11.5 sample with a loading/unloading rate of 0.71 s$^{-1}$. (b) The temperature-time curve corresponds to the overall average marked by the black area in Figure. S11c. The inset is an enlarged view of the temperature time curve of the loading process. (c) IR images of the corresponding points in the temperature-time curve in Figure. S6b.

**Table S3.** Comparison on the maximum stress and $\Delta T_{ad}$ in the eCE cycle between the $Mn_{50}Ni_{37.5}Ti_{12.5}$ alloy and other elastocaloric materials reported in the literatures.

| Alloy | Sample status | Max stress | $|\Delta S|$ (J kg$^{-1}$ K$^{-1}$) | Loading / Unloading $|\Delta T_{ad}|$ (K) | Loading / Unloading Strain rate (s$^{-1}$) | Refs |
|---|---|---|---|---|---|---|
| **NiMn-based** | | | | | | |
| $Mn_{50}Ni_{37.5}Ti_{12.5}$ | Polycrystal | 1527 | 63.3 | 57.2 / 6.6 | 1.07 / 1.07 | This work |
| $Ni_{50}Mn_{32}Ti_{18}$ | Polycrystal | 980 | 52.9 | 10.7 / 10.5 | 0.028 / 0.028 | [1] |
| $Ni_{49}Mn_{33}Ti_{18}$ | Oriented | 860 | 51.0 | 33.0 / 37.3 | 1.7 / 3.4 | [9] |
| $Ni_{50}Mn_{31.6}Ti_{18.4}$ | Single crystal | 800* | - | 10.0 / 24.8 | 0.001 / 2.0 | [10] |
| $Ni_{35.5}Co_{14.5}Mn_{35}Ti_{15}$ | Oriented | 180 | 42.1 | 11.5 / 5.8 | 0.021 / 0.021 | [11] |
| $(Ni_{50}Mn_{31.5}Ti_{18.5})_{99.8}B_{0.2}$ | Polycrystal | 700 | 45.0 | 26.9 /31.5 | 0.16 / 5.33 | [12] |
| $Mn_{50}Ni_{32}Sn_7Co_{11}$ | Polycrystal | 1162* | - | - / 2.6 | slow / 0.021 | [13] |
| $Ni_{45}Mn_{44}Sn_{11}$ | Polycrystal | 519* | 42.0 | - / 11.0 | slow / 0.048 | [14] |
| $Ni_{50}Mn_{35}In_{15}$ | Oriented | 400 | - | 11.5 / 19.7 | 0.015 / 1.4 | [15] |
| $Ni_{45}Mn_{36.5}In_{13.5}Co_5$ | Oriented | 260 | 12.2# | 8.6 / 5.8 | 0.0566 / 0.0566 | [8] |
| $Ni_{57}Mn_{18}Ga_{21.45}In_{3.55}$ | Oriented | 95 | 25.0 | - / 7.9 | - / 0.02 | [16] |
| $Ni_{50}Mn_{28.5}Cu_{4.5}In_{14}Ga_3$ | Oriented | 1287* | 29.3 | 3.4 / 19.0 | 0.002 / 2.0 | [17] |
| **NiTi** | | | | | | |
| $Ni_{48.9}Ti_{51.1}$ | Wire | 1103 | 35.1# | 25.0 / 21.0 | 0.2 / 0.2 | [7] |
| $Ni_{55.9}Ti_{44.1}$ | Tubes | 1180 | - | 27.0 / 20.0 | 0.068 / 0.068 | [18] |
| $Ni_{50.9}Ti_{48.9}Si_{0.2}$ | Ribbon | 600 | 40.9# | - / 11.3 | - / 0.15 | [19] |
| $Ti_{52.8}Ni_{22.2}Cu_{22.5}Co_{2.5}$ | Film | 690 | 21.8 | 8.9 / 11.2 | 0.25 / 0.25 | [20] |
| $Ti_{50}Ni_{44}Cu_5Al_1$ | Nanocrystalline | 750 | 31.8 | - / 17.4 | - / 2.0 | [21] |
| **Cu-based** | | | | | | |
| $Cu_{71.5}Al_{17.5}Mn_{11}$ | Oriented | 215 | 25.0 | 8.0 / 12.8 | 0..05 / 0.13 | [22] |
| $Cu_{70.4}Al_{17.2}Mn_{12.4}$ | Single crystal | 810 | 24.0 | - / 12.3 | - / 0.0011 | [23] |
| $Cu_{71.1}Al_{17.2}Mn_{11.7}$ | Wire | 400 | - | 12.6 / 11.9 | 0.15 / 0.15 | [24] |
| $Cu_{59.1}Zn_{27}Al_{13.8}Zr_{0.1}$ | Oriented | 550 | 23.6 | - / 14.2 | - / 0.2 | [25] |

* compressive strength; b) calculated by indirect method.